\documentclass[twocolumn,aps,prl,showpacs]{revtex4}
\usepackage{amsmath,graphicx,color,amssymb}

\begin{document}
\title{Band-like motion and mobility saturation in 
organic molecular
  semiconductors}

\author{S. Fratini$^{1,2}$ and S. Ciuchi$^{3}$} 

\affiliation{$^1$Institut N\'eel - CNRS \& Universit\'e Joseph Fourier 
BP 166, F-38042 Grenoble Cedex 9, France\\ 
$^2$Instituto de Ciencia de Materiales de Madrid, CSIC, Sor
  Juana In\'es de la Cruz 3, E-28049 Madrid, Spain\\ 
$^3$SMC-INFM Research Center, CNISM and 
Dipartimento di Fisica\\ 
Universit\`a dell'Aquila, 
via Vetoio, I-67010 Coppito-L'Aquila, Italy}

\date{\today}

\begin{abstract}  
We analyze a model that accounts for the inherently
large thermal lattice fluctuations associated to the weak van der Waals 
inter-molecular bonding in crystalline organic semiconductors.
In these materials 
the charge mobility generally exhibits a
``metallic-like'' power-law behavior, with no 
sign of thermally activated hopping characteristic of carrier 
self-localization, despite apparent 
mean-free-paths comparable or lower than the inter-molecular spacing.  
Our results show that such puzzling transport regime 
can be understood from 
the simultaneous presence of band carriers and incoherent 
states that are dynamically localized by the thermal lattice disorder.  
\end{abstract} 

\maketitle

In the past years, the development of crystal growing techniques 
has led to the production of molecular
organic semiconductors with extremely low structural disorder, attaining an 
upper intrinsic limit of the carrier mobility $\mu \sim 10^1-10^2 cm^2/Vs$, 
whose explanation still challenges the scientific community.
\cite{GershensonRMP06}
Surprisingly,  despite decades of active investigation 
there is still no well-established theory of charge transport in these
materials.
Attempts to generalize the usual concepts that successfully apply to 
inorganic semiconductors 
have revealed unsatisfactory, failing to provide 
a complete understanding of the fundamental mechanisms governing the 
charge mobility.  
In this respect, there has been historically a duality 
between a conventional 
band-like description\cite{Glarum63,Friedman65}
relying on the existence of well defined ``Bloch'' states 
and the opposite view of self-localized (polaronic) 
carriers hopping incoherently 
from molecule to molecule,\cite{GosarChoi66,
MunnSilbey85,Kenkre89,silinshreview95}
but no conclusive agreement has been reached
if either of these limiting descriptions applies in practice. 
In fact, the most puzzling feature of charge transport in 
crystalline organic semiconductors is that they generally
exhibit a ``band-like'' mobility characterized by a power-law
decrease with temperature,\cite{Karl85,GershensonRMP06} 
but with absolute values close to or even below the Mott-Ioffe-Regel
limit around room temperature, i.e. with 
apparent mean-free-paths that are
comparable or even lower than  the inter-molecular spacing.\cite{Cheng03} 
This observation clearly suggests a breakdown of
conventional band behavior and has been often taken as an indication
of polaronic carrier localization, even though the absence
of a distinctive thermally activated mobility is at odds with this 
interpretation. Recently, an alternative mechanism has been suggested
where  the electronic wavefunction is localized
 by the large thermal lattice fluctuations\cite{Troisi,Picon07} 
rather than by polaronic self-trapping effects, a picture
that is indeed compatible with the observed mobilities, $d\mu/dT<0$.
However, 
new experimental evidence on the existence of 
band-like carriers has gathered 
from different techniques such as transient
photoconductivity,\cite{Ostroverkhova,Moses} optical absorption 
\cite{Fischer,Basov} and photoemission \cite{ARPES}, 
which has  once more reopened the debate.

In this Letter, we revisit the traditional duality of band-like
{\it vs.} localized carriers in
highly conducting crystalline organic semiconductors,  
where electron-lattice interactions are too weak to induce polaronic 
self-localization. In these systems 
the limiting paradigm of thermally activated hopping
is replaced by a different high temperature regime 
where the carriers diffuse incoherently in the presence of large 
thermal lattice fluctuations,  giving rise to a weakly metallic
behavior which is analogous to the ``resistivity saturation'' 
characteristic of bad metals. \cite{Millis,GunnarssonNat}
In the experimentally accessible 
temperature range, our results show that   
{\it both} band-like features characteristic of Bloch electrons 
and incoherent states of a more localized nature
 are simultaneously present in the single-particle
excitation spectrum, 
albeit {\it at different energy scales}. 
Accordingly, different experimental probes will see alternatively one 
feature or the other, which is likely at the origin of
the long-standing controversy on the microscopic 
identity of the charge carriers in these materials. 
Because of this duality, the charge dynamics cannot be 
understood in terms of a single microscopic mechanism, since  
band-like carriers as well as incoherent excitations 
of a more localized nature both
contribute to the electrical mobility.

In  organic  semiconductors the molecules  
are bound together by weak van der Waals
forces. These are at the origin of the most salient characteristics of
these materials, that are held responsible for their peculiar 
electronic properties: the presence of extremely narrow electronic
bandwidths,  comparable to the thermal energy at room temperature, and
their mechanical softness,  reflected in 
large fluctuations of the molecular lattice around its equilibrium
structure.
The thermal lattice vibrations give rise to a large (dynamic) disorder in the 
inter-molecular transfer integrals $t$
which is currently identified as the dominant intrinsic limiting
factor of the mobility in these compounds.
\cite{MunnSilbey85,Hannewald,Troisi,TroisiAdv,Laarhoven} 
Other microscopic mechanisms, such as the  reorganization of 
electronic states within the 
molecules (the intra-molecular electron-phonon interaction), 
have been shown to have a weaker
effect on charge transport,\cite{Troisi,TroisiAdv} especially
in crystals of large molecules such as pentacene
and rubrene where the highest mobilities are currently achieved  
\cite{Devos,Coropceanu02}.
The above features can be formalized in the following model
Hamiltonian\cite{Glarum63,Friedman65,MadhukarPost77,MunnSilbey85,Troisi}
\begin{eqnarray}
  H&=& - t\sum_i f (X_i-X_{i+\delta}) \; (c^+_i  c_{i+\delta} + c^+_{i+\delta}  c_{i}) \nonumber\\
  &+&\sum_i \frac{M \omega_0^2}{2} X_i^2
  \label{eq:SSH}
\end{eqnarray}
where $c^+_i$ ($c_i$) are creation (annihilation) operators for an 
electron on a given molecule and the function $f$
measures the variation of the transfer integrals between 
neighboring molecules (at sites $i$ and $i+\delta$) 
due to their relative displacements, $X_i-X_{i+\delta}$. In the
following we shall assume  for clarity 
a linear dependence $f(x)=1-\alpha x$, although our method can be easily generalized
 to other forms of $f(x)$. \cite{linear}

To solve the model Eq. (\ref{eq:SSH}) 
we treat the $X_i$  as classical variables, 
assuming that the lattice dynamics are slower than any other time-scale
in the system. This is justified in virtue of the extremely 
low frequencies of the 
inter-molecular phonons that couple to the electronic motion, 
smaller than both the thermal and the band energy scale 
($\hbar\omega_0\simeq 4-9meV$, $t\simeq 130meV$
in rubrene\cite{TroisiAdv}, see \cite{Troisi,Hannewald,WangJCP07}
for different compounds). In this case the electronic properties 
depend on a single dimensionless coupling parameter 
$\lambda=\alpha^2t/(2M\omega_0^2)$.
We evaluate the non-local  electron Green's function
$G(i,j,\omega,X)$ ($i,j=$ lattice sites) 
for a statistical set of lattice configurations $X=\{X_i\}$ 
corresponding to a random extraction of the molecular displacements
$X_i$ out of a gaussian distribution $P(X_i)\propto
\exp(-M\omega_0^2X_i^2/2k_B T)$.    
The electronic problem at each given $X$ is solved
numerically  using an algorithm based on
regularization of recursion formulas \cite{Yamani} as an
alternative to common exact diagonalization techniques, so that system
sizes up to $N=2^{16}$ sites can be achieved. 
In order to control the residual finite-size effects  we attach
infinite metallic leads to both ends of the linear chain. 
The dynamical nature of the lattice disorder is restored
upon averaging over up to $10^6$ lattice configurations $X$,
defining the physically observable Green's function 
 as $G(|i-j|,\omega)=\langle G(i,j,\omega,X) \rangle_X$.
The   electrical conductivity is then evaluated via the Kubo formula
expressed in terms of the exact electron propagators, neglecting vertex
corrections as was done in Ref. \cite{Millis} and validated in Ref. \cite{GunnarssonNat} in a model for metallic fullerenes. 
This procedure allows to establish a direct connection between the excitation
spectrum and the transport properties, capturing 
the essential aspects of the transport mechanism 
which stem directly from the dual nature of the  
electronic states (see below).
\cite{note}
Here we present results obtained
on a one-dimensional
molecular chain, ideally corresponding to the direction of highest mobility in
the organic crystal.

\begin{figure}
\centering
\includegraphics[width=\columnwidth]{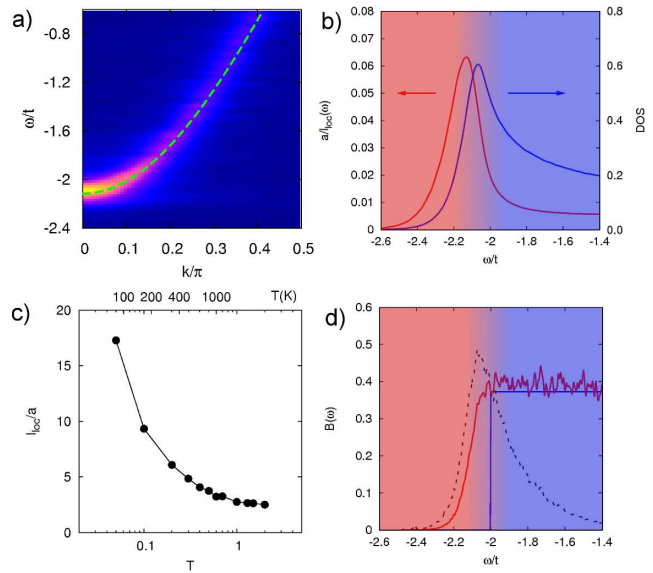}
  \caption{Quantities calculated from the solution of the
    model Eq.(1), representative for crystalline rubrene at room temperature 
($T=0.2t$ with $t=143meV$ and $\lambda=0.17$  as 
taken from Ref.\cite{TroisiAdv}).  
a) spectral function $A(k,\omega)$, showing a well defined band with a
weakly renormalized dispersion;
b) energy resolved localization length (red) and
density of states (blue); 
c) spatial extension of the electronic 
wave-function; 
d) current-current correlation function (red), and the same function 
multiplied by the thermal occupation factor as it appears in Eq. (2) 
(black dashed).
  The blue line is the Boltzmann result for band electrons, 
  $B(\omega)=\sum_k \delta(\omega-\epsilon_k) v_k^2
  \tau_k$, with $\epsilon_k$ the band dispersion and 
  $\tau_k$ the quasiparticle lifetime.
  The data  are evaluated with $10^6$
  statistical samples, respectively on 
  $a),b) N=2^{10}, c),d) 2^{8}$  lattice sites. 
}
  \label{fig:spect}
\end{figure}

Fig. \ref{fig:spect}a shows the single-particle spectral function
$A(k,\omega)=-Im \ \tilde G(k,\omega)/\pi$ [$\tilde G(k,\omega)$ being the Fourier
transform of the Green's function $G(|i-j|,\omega)$]  that carries information on the energy dispersion and lifetime of the extended Bloch waves. 
Owing to the relatively low values of 
$\lambda=0.05-0.2$ in organic
semiconductors 
\cite{Hannewald,WangJCP07,Troisi,TroisiAdv}, 
no polaronic self-trapping is expected. Indeed, at such moderate
coupling strengths 
one obtains a weakly
renormalized band dispersion (dashed line) and  
a quasi-particle scattering rate that increases with temperature, 
$1/\tau \propto
\lambda \langle (X_i-X_j)^2 \rangle \sim \lambda k_B T$, being
proportional to the disorder induced by the thermal lattice 
motion.
We see from Fig. \ref{fig:spect}a that even at room temperature the
scattering rate is sufficiently small compared to the bandwidth 
to allow for the existence of well-defined Bloch states. 
Moreover, the quasi-particles appear to be sharper near the band edge,
i.e.  precisely for those 
states that would be populated in a clean
non-degenerate semiconductor. 
The states of low-momentum
are partly protected by the ``off-diagonal'' nature of the
inter-molecular electron-lattice coupling, i.e. the fact that  
the dominant interaction acts on 
the transfer integrals 
rather than on the molecular energy levels. 
A perturbative calculation yields
$1/\tau_{k}\propto k$ at low momenta, but 
our numerical data show that the suppression of scattering is not
complete and  saturates to a finite value 
$1/\tau_{k=0}\simeq \pi\lambda k_BT$. 

While the $k$-space analysis carried out above
would  point to a 
conventional scenario based on weakly scattered momentum
states, a different conclusion is reached if one looks at physical
quantities in real space.
The idea is that if one takes a shapshot of the system at a given time 
the electronic wave-function will be apparently localized by the instantaneous
disordered landscape of lattice deformations, even though the carrier 
actually diffuses on the long time owing to the  dynamics of the
molecular lattice (diffusion  eventually  sets in at times longer than
the period of inter-molecular vibrations, which is itself much larger than
the timescale of electronic motion).
\cite{Troisi,Picon07} 
To illustrate this point 
we calculate the average spread of the 
electronic wave-function for the states at a given energy,  $l_{\rm
  loc}(\omega)$. Following 
Refs. \cite{Thouless,Theodorou}, 
this quantity can be
expressed in terms of the  density of states (DOS) $\rho(\omega)=G(0,\omega)$
 as 
\begin{equation}
{l_{loc}^{-1}(\omega)}=\int d\omega^\prime \rho(\omega) 
\rho(\omega^\prime)\log(|\omega-\omega^\prime|/|\omega^\prime|).
\end{equation}
The comparison of the localization length 
with the actual DOS
in Fig.\ref{fig:spect}b 
shows that two classes of states of very different nature
{\it coexist} in the excitation spectrum.
The states within the band, ($\omega\gtrsim -2t$), 
are effectively delocalized over hundreds of molecular sites
and can therefore be treated in the wave representation. However, 
the dynamical disorder induced in the inter-molecular transfer integrals by the
thermal lattice vibrations
gives rise to a tail of incoherent excitations emerging in the 
DOS below the band edge, ($\omega\lesssim -2t$).
These states have a much shorter spatial extent 
and are ultimately responsible for the localized character of the particles:
the actual localization length is obtained as the thermal average
$l_{loc}^{-1}\propto \int l_{loc}^{-1}(\omega) e^{-\omega/k_BT}$, 
which is dominated by such incoherent tail states where
$l_{loc}^{-1}(\omega)$ is maximum.  
As a result, real-space probes of the electronic properties\cite{Marumoto} 
will point to the presence of  particles localized on few molecular sites
 \cite{Troisi}, as shown in Fig. \ref{fig:spect}c. 
On the other hand, momentum-resolved probes such as
angle-resolved photoemission\cite{ARPES} should provide a picture 
in agreement with
the wave nature of electrons, as shown in Fig. \ref{fig:spect}a.

To ascertain how the duality evidenced in the 
single-particle excitation spectrum
is reflected in the
transport properties, one needs to go beyond the semi-classical 
Boltzmann treatment which is valid for band-like carriers alone. 
We therefore  resort to the Kubo formula
\begin{equation}
  \label{eq:mu}
  \mu=\frac{\pi\mu_0}{Zk_BT} \int
d\omega B(\omega) e^{-\omega/k_BT}
\end{equation} 
which expresses the mobility in terms of the current-current
correlation function in the limit of zero exchanged  momentum, 
$B(\omega)=\underset{q\to 0}{\lim}\int d\omega e^{i\omega t} 
\langle [J(t),J(0)] \rangle$ ($\mu_0= ea^2/\hbar$ carries the dimensions of
mobility, with $a$ the average inter-molecular spacing, and $Z=\int d\omega 
\rho(\omega) e^{-\omega/k_BT}$). 
Inspection of Eq. (\ref{eq:mu}) shows
that the temperature dependent function $B(\omega)$
represents the contribution to the
electrical conduction from the states at energy $\omega$, which can be
viewed as an ``energy-resolved'' carrier mobility.
The current-current correlation function 
calculated from the exact
electron propagators clearly shows that conduction from 
the band-like carriers seen in the spectral function of Fig. 
\ref{fig:spect}a is reasonably well
described by Boltzmann theory, that predicts a constant $B(\omega)$ 
within the band, see Fig. \ref{fig:spect}d.
However, 
an additional transport channel emerges due to the states below the band edge, 
corresponding to the incoherent tail seen 
in the DOS of Fig. \ref{fig:spect}b. Both band-like and incoherent states
are therefore expected to  contribute to the transport mechanism.
Their relative role will depend crucially on  the temperature,  
being essentially 
controlled by the amount of thermal lattice disorder (which sets
the size of the incoherent tail)
and to a lesser extent 
by their respective thermal population, via the exponential term in
Eq.(\ref{eq:mu}). This interplay is best visualized by looking
directly at the integrand of Eq. (\ref{eq:mu}), i.e.
the bubble $B(\omega)$ weighted by the Boltzmann factor 
(black dashed curve in \ref{fig:spect}d): at $T=0.2t$  
most of the weight is located right across the energy scale that
separates the dynamically localized from the delocalized states.
Upon increasing the temperature, 
the relative weight of the incoherent states
progressively increases up to the point  
where band states get competely washed out, as their
lifetime becomes shorter than the inter-molecular transfer time
$\tau\lesssim t^{-1}$.

\begin{figure}
 \centering
  \includegraphics[width=7cm]{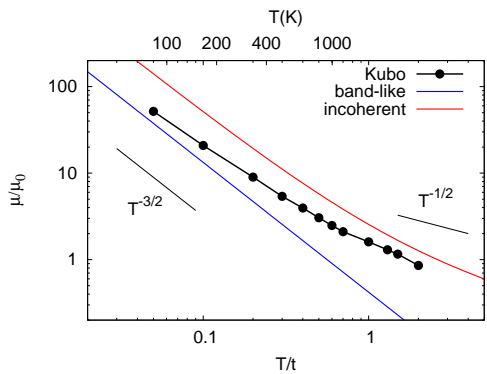}
  \caption{Mobility calculated from the
    Kubo formula in terms of the electron propagators obtained
    from the numerical simulation. For realistic
    bandwidth and electron-lattice interaction 
   parameters for  organic semiconductors (see Fig. 1), both band-like
   conduction and incoherent diffusion contribute to the transport
   mechanism 
   in the relevant temperature range around room temperature. 
}
  \label{fig:mob}
\end{figure}

The mobility obtained from Eq. (\ref{eq:mu})
 is illustrated in Fig. \ref{fig:mob}
 together with the limiting results valid for band
electrons alone  ($\mu \sim T^{-3/2}$ and $\sim T^{-2}$  
are obtained from the Boltzmann treatment 
 at $k_BT\lesssim 2t$ and
$k_BT\gtrsim 2t$, see Refs. \cite{Glarum63,Friedman65}) 
and for the random
diffusion of fully incoherent states\cite{GosarChoi66,SumiJCP79}.
The latter  corresponds to the regime of
``mobility saturation'' \cite{Millis,GunnarssonNat,Calandra02} 
 that sets in when the
the  mean-free path for band
electrons falls below the inter-molecular spacing, $l_{\rm mfp}\lesssim
a$, leading to a complete loss of momentum conservation. 
Above this temperature, which essentially coincides with the condition
$\tau\lesssim t^{-1}$ given above, the charge transport proceeds
via the  incoherent diffusion of carriers being randomized
at each jump as in a classical random walk. This mechanism
results in a rather flat ``metallic-like''
power-law behavior\cite{GosarChoi66,SumiJCP79},
$\mu\sim T^{-1/2}$, that should be contrasted
with the exponentially activated behaviour characteristic of 
self-trapped polarons, occurring for much larger values
of the electron-lattice interaction, $\lambda \gtrsim 1$.
We see from Fig. \ref{fig:mob} that the calculated mobility  
undergoes a very broad crossover where it 
progressively interpolates between the band-like regime 
and the incoherent diffusion regime. 
In the temperature range of interest, these two competing  
mechanisms 
are simultaneously present and
combine as independent transport channels.
\cite{Calandra02,Fratini03}

Our results based on a model that incorporates the essential
ingredients relevant for crystalline organic semiconductors
show that neither of the two limiting pictures
proposed in the past 
adequately describes the charge dynamics. Firstly, 
polaronic self-localization is prevented in high-mobility organic
semiconductors  by the relatively weak electron-lattice interactions.
However, the opposite view of band-like transport is not appropriate either: 
due to the large thermal lattice fluctuations arising from the
mechanical softness of these systems,
incoherent states having a localized nature emerge and are found to coexist 
with more conventional Bloch states. 
Both band-like and incoherent states 
actually 
contribute to the 
electronic properties of these materials but are expected to show up 
differently according to the experimental probe.
Also, depending on the actual values of the 
bandwidth and inter-molecular electron-lattice coupling 
relevant to a given material, a whole range of intermediate behaviours
between the two limiting transport scenarios can arise,
which might explain the large variability of 
exponents of the mobility $\mu\sim T^{-n}$ observed in the
literature.\cite{Karl85,GershensonRMP06}

S. F. acknowledges
financial support from MICINN (Consolider CSD2007-00010) and from
the the Comunidad de Madrid (CITECNOMIK).
S. C. aknowledges the hospitality of ICMM-CSIC.


\begin{thebibliography}{30}

\bibitem{GershensonRMP06} M. E. Gershenson, V. Podzorov and
A. F. Morpurgo.  Rev. Mod. Phys. 78, 973 (2006)
\bibitem{Glarum63}  S. H. Glarum,  J. Phys. Chem. Solids  24, 1577 (1963)
\bibitem{Friedman65} L. Friedman,  Phys. Rev. 140, A1649 (1965)
\bibitem{GosarChoi66}  P. Gosar \&  S.-I.  Choi,  Phys. Rev. 150, 529 (1966)
\bibitem{MunnSilbey85}  R. W. Munn  \&   R.  J. Silbey, 
Chem. Phys. 83, 1854 (1985) 
\bibitem{Kenkre89} V. M. Kenkre, J. D. Andersen, D. H. Dunlap, and C. B. Duke
Phys. Rev. Lett. 62, 1165 (1989).
\bibitem{silinshreview95} E. A. Silinsh, A. Klimkans, S. Larsson, V. Capek. 
 Chem. Phys. 198, 311 (1995) 
\bibitem{Karl85} N. Karl, {\it Organic Electronic Materials}, ed. R.
Farchioni \& G. Grosso, Springer-Verlag, Berlin, pp. 283-326 (2001)
\bibitem{Cheng03}  Y. C. Cheng {\it et al.},  
J. Chem. Phys 118, 3764 (2003) 

\bibitem{Troisi} A. Troisi \& G. Orlandi,  Phys. Rev. Lett. 96, 086601 (2006)

\bibitem{Picon07} J. D. Picon, M. N.  Bussac, L. Zuppiroli, 
 Phys. Rev. B 75, 235106 (2007)
\bibitem{Ostroverkhova} 
O. Ostroverkhova {\it et al.}, 
 Appl. Phys. Lett. 88, 162101 (2006)
\bibitem{Moses} D. Moses,
 C. Soci, X. Chi, and A. P. Ramirez.  
Phys. Rev. Lett. 97, 067401 (2006)
\bibitem{Fischer} M. Fischer, M. Dressel, B. Gompf
A. K. Tripathi and J. Pflaum.  Appl. Phys. Lett. 
89, 182103 (2006)
\bibitem{Basov} Z. Q. Li {\it et al.}, 
Phys. Rev. Lett. 99, 016403 (2007)
\bibitem{ARPES} H. Kakuta {\it et al.}, 
Phys. Rev. Lett. 98,  247601 (2007)


\bibitem{Millis} A. J. Millis, J. Hu , S.  Das Sarma,  
{\it Phys. Rev. Lett.} {\bf 82}, 2354 (1999)

\bibitem{GunnarssonNat} O. Gunnarsson  \&   J.E. Han,   {\it Nature}
  {\bf 405}, 1027 (2000)  

\bibitem{Hannewald} K. Hannewald \& P. A. Bobbert. 
 Phys. Rev. B 69, 075212 (2004).

\bibitem{TroisiAdv} A. Troisi,   Adv. Mat. 19, 2000 (2007)

\bibitem{Laarhoven} H. A. Laarhoven et al., J. Chem. Phys. 129, 044704 (2008)


\bibitem{MadhukarPost77} A. Madhukar \& W. Post,  Phys. Rev. Lett. 39, 1424 (1977)



\bibitem{SumiJCP79} H.   J. Sumi, Chem. Phys. (1979)

\bibitem{WangJCP07} L. J. Wang, Q. Peng, Q. K. Li, and Z. Shuai.
 J. Chem. Phys. 127, 044506 (2007)



\bibitem{Devos} A. Devos \& M. Lannoo, 
 Phys. Rev. B 58,
8236 (1998)

\bibitem{Coropceanu02} V. Coropceanu {\it et al.}, 
 Phys. Rev. Lett. 89, 275503 (2002)
\bibitem{linear}
{The validity of the linear
approximation can be justified  by observing that in the materials
under study the statistical distribution of transfer integrals  
essentially follows the gaussian
shape determined by the thermal distribution of lattice displacements (see e.g. Fig.1a of Ref. \cite{TroisiAdv}),
while more general non-linear couplings would lead to markedly asymmetric and
non-gaussian distributions. }


\bibitem{Yamani}
 H. A. Yamani \&  M. S. Abdelmonem,  
J. Phys. A: Math. Gen.,  30, 2889 (1997)



\bibitem{Thouless}  D. J. Thouless,  J. Phys. C 5, 77 (1972) 
\bibitem{Theodorou} G. Theodorou \&  M. H. Cohen,  Phys. Rev. B
  13, 4597 (1976)



\bibitem{note} Dropping  vertex corrections is physically meaningful
  as it discards precisely those quantum interference effects
  that would lead to genuine electron localization in the presence of static
disorder, but are absent for the dynamical
lattice disorder considered here, see e.g. [P. A. Lee \&
T. V. Ramakrishnan, Rev. Mod. Phys. 57, 287 (1985)]


\bibitem{Marumoto} K. Marumoto, S. I.  Kuroda, T. Takenobu, Y. Iwasa,
  Phys. Rev. Lett. 97, 256603 (2006)

\bibitem{Calandra02} 
M. Calandra \& O. Gunnarsson,
Phys. Rev. B 66, 205105 (2002)

\bibitem{Fratini03} S. Fratini \& S.  Ciuchi, 
 Phys. Rev. Lett. 91, 256403 (2003)


\end{thebibliography}
\end{document}